\documentclass[11pt,aps,prd,groupedaddress,preprintnumbers,a4paper]{revtex4}
\usepackage[utf8]{inputenc}
\usepackage[english]{babel}
\usepackage{amsmath}
\usepackage{amsfonts}
\usepackage{subfig}
\usepackage{amssymb}
\usepackage{graphicx}
\usepackage{epstopdf}
\usepackage{float}
\usepackage[margin=3cm]{geometry}
\usepackage{setspace}
\usepackage{caption}

\begin{document}
\preprint{FTUV-10-29}
\preprint{IFIC/16-73}

\title{A composite tale of two right-handed neutrinos}
\author{Gabriela~Barenboim\footnote{Gabriela.Barenboim@uv.es}}
\affiliation{Departament de F\'{\i}sica Te\`orica and IFIC, Universitat de 
Val\`encia-CSIC, E-46100, Burjassot, Spain.}
\author{Cristian~Bosch\footnote{Cristian.Bosch@uv.es}}
\affiliation{Departament de F\'{\i}sica Te\`orica and IFIC, Universitat de 
Val\`encia-CSIC, E-46100, Burjassot, Spain.}


\begin{abstract}
In this work, we develop a model for Higgs-like composites based on two generations of right handed neutrinos which condense. We analyze the Spontaneous Symmetry Breaking of the theory with two explicit breakings, setting the different scales of the model and obtaining massive bosons as a result. Finally, we calculate the gravitational wave imprint left by the phase transition associated to the symmetry breaking of a generic potential dictated by the symmetries of the composites.
\end{abstract}
\maketitle

\section{Introduction}
The observation of neutrino oscillations suggests that there must be physics beyond the Standard Model of particle physics due to the inclusion of right-handed neutrinos that are sterile with respect to the standard model gauge interactions.  These additional states mix with the usual neutrinos of the Standard Model through Yukawa interactions with the left-handed leptons.  If these new sterile neutrinos become massive, then masses for the Standard Model neutrinos are also generated by the mixing between the sterile and active neutrinos. It is important to notice that the right-handed mass scale is not protected by any symmetry and therefore can take any value. On general grounds, one can assume it will take the highest possible value.

Despite not being active players of the SM dynamics, as mentioned before, right-handed (RH) neutrino existence is well-motivated by the evidence of non-zero neutrino masses provided by the oscillation experiments and by the yet unconfirmed but still intriguing evidence of reactor and accelerator experiments for additional light sterile neutrino states. Without any direct experimental evidence, there could be a wide range (from sub-eV to $10^{15}$ GeV) for the scale of masses for the right-handed neutrinos.

While Majorana mass terms for the sterile neutrinos could be directly added to the theory, we wish to explore a dynamical origin of these masses through sterile neutrino condensates.  Inspired by the BHL \cite{Bardeen:1989ds} model of composite Higgs fields, the role of the top quark condensates will be played by a condensate of the sterile neutrinos.  The sterile neutrinos have no SM gauge interactions except for the gravitational interactions which are flavor neutral.  We will assume that gravity or some other flavor neutral dynamics is responsible for driving the formation of the sterile neutrino condensates. The minimal model of sterile neutrino condensates will require at least two flavors of sterile neutrinos to explain the observed neutrinos oscillations.

With two flavors of sterile neutrinos and flavor neutral dynamics, the neutrino condensates will transform as a symmetric tensor representation of the U(2) sterile neutrino flavor symmetry.  The dynamical breaking of this flavor symmetry is triggered by the formation of the sterile neutrino condensates and a number of composite scalars that will have their own low-scale interactions.  A possible realization of the dynamics involved in the condensation can be reviewed in \cite{Barenboim:2010db} where the gravitational interactions of the RH sterile neutrinos is the mechanism inducing the condensation.  This gravitational attraction, or different and undetermined short-range interactions, is independent of the sterile neutrino flavor thus forming composite scalars that are not constrained by the need for neutralizing their flavor charge implying a larger number of scalar bound states.  While there could be a larger number of sterile neutrinos, we will focus on the minimal set needed to explain the observed neutrino oscillations which consists of the two flavors mentioned above.  The two-flavor model will generate masses for only two of the three SM neutrinos so one of the neutrino mass eigenstates will have zero mass.

In the BHL model, the Higgs field is formed as a top quark bound state at a high scale, $\Lambda$.  At a much lower scale, the electroweak symmetry is dynamically broken generating particle masses and a physical Higgs boson.  We will explore the nature of flavor symmetry breaking in the corresponding version of sterile neutrino condensate model.

The reader may wonder what is the interest, if there is any, of an effective theory that will have to deal with the appearance of new Nambu-Goldstone bosons and pseudo-Goldstone bosons once its symmetries are broken, and where to fit their masses in a particle catalogue that is not growing experimentally. However, this is not as troublesome as it may seem, and we can explore the convenience of these condensates for the treatment of some theoretical puzzles still unsolved.

One of them would be Cosmological Inflation. Keeping in mind  models based on the same idea we are dealing with here \cite{Barenboim:2008ds}, we will propose a series of bosons after symmetry breaking that could be interesting for this purpose.

Another possibility would be to explore the prospect of solving the Electro-Weak (EW) vacuum instability present in the running of the Higgs quartic coupling \cite{EliasMiro:2012ay}. Unfortunately, as we will see, our particular realization of RH neutrino condensation will prove not enough to fix this situation. RH neutrino composite scalars are a promising line of work for solving this problem nonetheless.

Additionally, we will address the experimental signature of the phase transition we are proposing, its imprint in the gravitational wave spectrum. In February 2016, Advanced LIGO (Laser Interferometer Gravitational Observatory) \cite{grav-wave-ligo} confirmed experimentally the existence of gravitational waves. Being the scientifical outbreak of the year, it would be quite interesting to determine the signature that our phase transition would leave in the gravitational spectrum. A neutrino condensate would, thus, be recognizable if a gravitational wave signal matches the theoretical prediction of the phase transition corresponding to a spontaneous symmetry breaking.

This paper is organized as follows: Section \ref{SSB} is dedicated to introduce the scalar fields, the change into vectorial notation and calculating the Spontaneous Symmetry Breaking (SSB) of the scalar theory. Section \ref{explicit} introduces an explicit breaking in a lower energy scale and in section \ref{yukawa} we persent the last breaking, due to the Yukawa interaction of the scalar, being the lowest energy scale of the theory. In section \ref{grav}, the print left as gravitational waves by a potential featuring the same symmetries as the one presented in this analysis is studied. After that, the conclusions are presented in section \ref{conclusions}.

\section{Symmetric tensor formalism and Spontaneous Symmetry Breaking}\label{SSB}
At the energy scale $\Lambda$ or due to the finite temperature of the universe, the sterile neutrinos can be expected to condense due to the attractive interaction of gravity or some other short-range dynamics. At low energy, the neutrino biliniars effectively become dynamical composite scalars transforming as a symmetric tensor representation of the U(2) flavor symmetry of the sterile neutrinos.

The simplest version of this dynamics invokes an attractive four-fermion interaction between the sterile neutrinos. Following BHL, we can solve for the dynamics of the composite field representing the fermion bilinear operators, $\varphi_{ij}=\left\langle\nu_i\nu_j\right\rangle$, where the latin indices represent sterile neutrino flavor. In analogy to the BHL calculation, we can compute the effective potential for the scalar field $\varphi_{ij}$, generated by integrating out the sterile neutrinos at one loop. We obtain:

\begin{equation}\label{potential}
V\left(\varphi\right)=\frac{1}{2}\lambda\varphi^\dagger_{ij}\varphi_{jk}\varphi^\dagger_{kl}\varphi_{li}-\frac{1}{2}M^2\varphi^\dagger_{ij}\varphi_{ji}
\end{equation}

By tuning the parameters of the four-fermion theory, the scalar bilinear fields become dynamical with the usual kinetic terms. In \cite{Barenboim:2010db}, a calculation of the gravity triggered condensation is computed using a particular large N limit.

For the sake of clarity we will change this notation to a more compact one which will also prove to be more convenient. Since we only have three scalars, $\varphi_{11}$, $\varphi_{12}$ and $\varphi_{22}$, we will change the tensor of scalars into a complex vector formalism $\vec{\varphi}=\left(\varphi_1,\varphi_2,\varphi_3\right)$,  defined as:

\begin{equation}
\varphi_{ij}=\left(\vec{\sigma}\vec{\varphi}i\sigma_2\right)_{ij};\;\varphi^\dagger_{ij}=\left(-i\sigma_2\vec{\sigma}\vec{\varphi}^*\right)_{ij}
\end{equation}

Which leaves the previous potential as follows:
\begin{equation}\label{bhl-pot}\resizebox{\linewidth}{!}{$
\begin{aligned}
V=&~\frac{1}{2}\lambda\mathrm{Tr}\left[\left(-i\sigma_2\right)\left(\vec{\sigma}\vec{\varphi}^*\right)\left(\vec{\sigma}\vec{\varphi}\right)i\sigma_2\left(-i\sigma_2\right)\left(\vec{\sigma}\vec{\varphi}^*\right)\left(\vec{\sigma}\vec{\varphi}\right)i\sigma_2\right]-\frac{1}{2}M^2\mathrm{Tr}\left[\left(-i\sigma_2\right)\left(\vec{\sigma}\vec{\varphi}^*\right)\left(\vec{\sigma}\vec{\varphi}\right)i\sigma_2\right]=\\
=&~\frac{\lambda}{2}\mathrm{Tr}\left[\left(\vec{\sigma}\vec{\varphi}^*\right)\left(\vec{\sigma}\vec{\varphi}\right)\left(\vec{\sigma}\vec{\varphi}^*\right)\left(\vec{\sigma}\vec{\varphi}\right)\right]-\frac{M^2}{2}\mathrm{Tr}\left[	\left(\vec{\sigma}\vec{\varphi}^*\right)\left(\vec{\sigma}\vec{\varphi}\right)\right]=\\
=&~\lambda\left[2\left(\vec{\varphi}^*\vec{\varphi}\right)^2-\left(\vec{\varphi}^*\vec{\varphi}^*\right)\left(\vec{\varphi}\vec{\varphi}\right)\right]-M^2\left(\vec{\varphi}^*\vec{\varphi}\right)
\end{aligned}$}
\end{equation} 

Now, using that the field has an intuitive geometrical interpretation, we can use rotational symmetry to redefine the vector field, beginning with a redefinition of the global phase that would leave $\varphi_2$ real, a rotation that would make $\varphi_2$ vanish, and a second global phase shift that would give a real first component, i.e. $\varphi_1=S$, the second component would remain $\varphi_2=0$ and the third component $\varphi_3=R+iI$ would be a complex field. Replacing that in the previous equation, we have:

\begin{equation}
V=\lambda\left[2\left(R^2+I^2+S^2\right)^2-\left(R^2+S^2-I^2\right)^2-4R^2I^2\right]-M^2\left(R^2+I^2+S^2\right)
\end{equation}

Minimizing it, we arrive to the following gap equations:
\begin{equation}
\begin{aligned}
A:\,\, & 2R\left[2\lambda\left(R^2+I^2+S^2\right)-M^2\right]=0\\
B:\,\, & 2I\left[2\lambda\left(R^2+I^2+3S^2\right)-M^2\right]=0\\
C:\,\, & 2S\left[2\lambda\left(R^2+3I^2+S^2\right)-M^2\right]=0
\end{aligned}
\end{equation}

Their three solutions and vacuum energies are given by:

\begin{equation}\label{veveq}
\begin{aligned}
AB:\,\, & R,\,I\neq0\Rightarrow S=0,\,R^2+I^2=\frac{M^2}{2\lambda},\,E=-\frac{M^4}{4\lambda} \\
AC:\,\, & R,\,S\neq0\Rightarrow I=0,\,R^2+S^2=\frac{M^2}{2\lambda},\,E=-\frac{M^4}{4\lambda} \\
BC:\,\, & S,\,I\neq0\Rightarrow I^2=S^2,\,R=0,\,4I^2=4S^2=\frac{M^2}{2\lambda},\,E=-\frac{M^2}{8\lambda}
\end{aligned}
\end{equation}

Thus, we can see that AB and AC are possible ground states whilst BC, with a higher energy, is not.

The AB and AC solutions each preserve a U(1) symmetry, associated to lepton number conservation. Thus, the breaking is U(2)$\rightarrow$ U(1) and there should be three Goldstone bosons and three heavy states associated to it. In both cases we can use the residual U(1) symmetry to align the real $\varphi$ field to be the 3 direction with all other components vanishing. Hence, AB and AC are really the same solution for the spontaneous symmetry breaking.

We can compute the mass spectrum by expanding around the vacuum with $R=\upsilon$, $\upsilon^2=M^2/2\lambda$ and $I,\,S=0$.

In summary we are left with three massive states, the real longitudinal field and the imaginary transverse fields, with masses given by $m^2=2M^2$, and three massless Goldstone modes, which will remain this way in the absence of any additional symmetry breaking.

\section{Explicit symmetry breaking terms}\label{explicit}
The explicit breaking of the lepton number can be achieved by adding the following term to the effective potential presented in equation \ref{bhl-pot}:

\begin{equation}\label{additionalterm}
\Delta V_1=-\mu^2\left(\vec{\varphi}^*\right)^2-\mu^2\left(\vec{\varphi}\right)^2+2\mu^2\left(\vec{\varphi}^*\vec{\varphi}\right)
\end{equation}

This term does not shift $\left\langle\varphi\right\rangle$ but the Goldstone field associated with the imaginary longitudinal field becomes a massive pseudo-Goldstone boson. The other two fields, associated with the real transverse part, remain massless.

The Yukawa interactions can provide the explicit breaking needed to make the remaining two Goldstone bosons massive. This one-loop effective Yukawa interaction can be written as:

\begin{equation}\label{secondaddition}
\Delta V_2\propto\left[\varphi^*_{ij}\varphi_{kl}\left(Y^\dagger Y\right)_{ik}\left(Y^\dagger Y\right)_{ji}\right]=\left[\vec{\sigma}\vec{\varphi}^\dagger\left(Y^\dagger Y\right)\vec{\sigma}\vec{\varphi}\left(i\sigma_2\right)\left(Y^\dagger Y\right)^T\left(-i\sigma_2\right)\right]
\end{equation}

Since the factor involving the Yukawa coupling is hermitian, we can write it in the general form:

\begin{equation}
\left(Y^\dagger Y\right)=\vec{\alpha}\vec{\sigma}+\beta;\;\left(\vec{\alpha},\,\beta\right)\;\;\mathrm{real}
\end{equation}

Inserting this into the above expression, we get:

\begin{equation}
\begin{aligned}
\Delta V_2\propto&\cdot\mathrm{Tr}\left[\left(\vec{\sigma}\vec{\varphi}^\dagger\right)\left(\vec{\sigma}\cdot\vec{\alpha}+\beta\right)\left(\vec{\sigma}\vec{\varphi}\right)\left(-\vec{\sigma}\cdot\vec{\alpha}+\beta\right)\right] \\
=& 2\left[-2\left(\vec{\alpha}\vec{\varphi}^\dagger\right)\left(\vec{\alpha}\vec{\varphi}\right)+\left(\vec{\alpha}^2+\beta^2\right)\left(\vec{\varphi}^\dagger\vec{\varphi}\right)+2i\beta\vec{\alpha}\left(\vec{\varphi}\times\vec{\varphi}^\dagger\right)\right] \\ \\
\end{aligned}
\end{equation}

\section{Scalar Potential}\label{yukawa}
Now we want to find the last breaking of the model, which will define the smallest scale associated to it and give mass to the remaining Goldstone bosons. The source of this breaking will be the effective Yukawa term, as we stated in the previous section. For that purpose, we write the symmetric tensor file as a complex three vector field $\vec{\varphi}$.

We presented the explicit symmetry breaking in the last section. In equation \ref{additionalterm}, we introduced a term $\mu^2$, which breaks lepton number but still preserves rotational symmetry. After that, we added a second term to the potential in equation \ref{secondaddition}, with the parameter $\vec{\alpha}$ representing the symmetry breaking due to Yukawa interactions. Thus, the full potential will be of the form:

\begin{equation}
\begin{aligned}
V=&\lambda\left[2\left(\vec{\varphi}\vec{\varphi}\right)^2-\left(\vec{\varphi}^*\right)^2\left(\vec{\varphi}\right)^2\right]-M^2\left(\vec{\varphi}^*\vec{\varphi}\right)-\mu^2\left(\vec{\varphi}^*\right)^2-\mu^2\left(\vec{\varphi}\right)^2 \\&+2\mu^2\left(\vec{\varphi}^*\vec{\varphi}\right)-\left(\vec{\alpha}\vec{\varphi}^*\right)\left(\vec{\alpha}\vec{\varphi}\right)+\vec{\alpha}^2\left(\vec{\varphi}^*\vec{\varphi}\right)\\ \\
\end{aligned}
\end{equation}

This potential has a minimum where its v.e.v. $\left\langle\vec{\varphi}\right\rangle$ is in the real part of $\vec{\varphi}$, pointing in the direction of $\vec{\alpha}$ (defined by the Yukawa interactions):

\begin{equation}
\vec{\varphi}=\vec{\varphi}^\dagger=\left\langle\vec{\varphi}\right\rangle\vec{\alpha}
\end{equation}

\begin{equation}
V=\lambda\left[2\left\langle\vec{\varphi}\right\rangle^4-\left\langle\vec{\varphi}\right\rangle^4\right]-M^2\left\langle\vec{\varphi}\right\rangle^2=\lambda\left(\left\langle\vec{\varphi}\right\rangle^2\right)^2-M^2\left\langle\vec{\varphi}\right\rangle^2
\end{equation}
\\
At its minimum: $\left\langle\vec{\varphi}\right\rangle^2=M^2/2\lambda$

Now expanding to second order in $\varphi$ fields, the transverse fields part of the potential will look like:
\begin{equation}
\begin{aligned}
V=&\lambda\left[4\left\langle\vec{\varphi}\right\rangle^2\left(\vec{\varphi}^\dagger_T\vec{\varphi}_T\right)-\left\langle\vec{\varphi}\right\rangle^2\left(\vec{\varphi}^\dagger_T\right)^2-\left\langle\vec{\varphi}\right\rangle^2\left(\vec{\varphi}_T\right)^2\right]-M^2\left(\vec{\varphi}^\dagger_T\vec{\varphi}_T\right)+\\&+\vec{\alpha}^2\left(\vec{\varphi}^\dagger_T\vec{\varphi}_T\right)-\mu^2\left(\vec{\varphi}^\dagger_T\right)^2-\mu^2\left(\vec{\varphi}_T\right)^2+2\mu^2\left(\vec{\varphi}^\dagger_T\vec{\varphi}_T\right)^2=\\
=&-\lambda\left\langle\vec{\varphi}\right\rangle^2\left(\vec{\varphi}^\dagger_T-\vec{\varphi}_T\right)^2+\vec{\alpha}^2\left(\vec{\varphi}^\dagger_T\vec{\varphi}_T\right)-\mu^2\left(\vec{\varphi}^\dagger_T-\vec{\varphi}_T\right)^2=\\
=&\vec{\alpha}^2\left(\mathrm{Re}\vec{\varphi}_T\right)^2+\left(\vec{\alpha}^2+4\lambda\left\langle\vec{\varphi}\right\rangle^2+4\mu^2\right)\left(\mathrm{Im}\vec{\varphi}_T\right)^2
\end{aligned}
\end{equation}

While the masses of the pseudo-Goldstone bosons arise as follows: for the real transverse field, the mass will be $m^2_{RT}=\vec{\alpha}^2$, whilst for the imaginary transverse field, it will have a value of $m^2_{IT}=\vec{\alpha}^2+4\lambda\left\langle\vec{\varphi}\right\rangle^2+4\mu^2=2M^2+4\mu^2+\vec{\alpha}^2$

Focusing now in the longitudinal fields part of the potential:
\begin{equation}
\begin{aligned}
V=&\lambda\left(\varphi^\dagger_L\varphi_L\right)^2-M^2\left(\varphi^\dagger_L\varphi_L\right)-\mu^2\left(\left(\varphi^\dagger_L\right)^2+\left(\varphi_L\right)^2-2\left(\varphi^\dagger_L\varphi_L\right)\right)=\\
=&\lambda\left(\left\langle\vec{\varphi}\right\rangle^2+2\left\langle\vec{\varphi}\right\rangle\varphi_{RL}+\varphi^2_{RL}+\varphi^2_{IL}\right)^2-M^2\left(\left\langle\vec{\varphi}\right\rangle^2+2\left\langle\vec{\varphi}\right\rangle\varphi_{RL}+\right.\\ &+\left.\varphi^2_{RL}+\varphi^2_{IL}\right)+4\mu^2\varphi^2_{IL}=\\
=&\left(6\lambda\left\langle\vec{\varphi}\right\rangle^2-M^2\right)\varphi^2_{RL}+\left(2\lambda\left\langle\vec{\varphi}\right\rangle^2-M^2\right)\varphi^2_{IL}+4\mu^2\varphi^2_{IL}
\end{aligned}
\end{equation}

For which the following masses are obtained:

\begin{equation}
\begin{aligned}
m^2_{RL}=&4\lambda\left\langle\vec{\varphi}\right\rangle^2=2M^2\\
m^2_{IL}=&4\mu^2 \\ \\
\end{aligned}
\end{equation}

We must note that there are three states associated with the high mass scale $M$, one state $\mu$ associated with the lepton number symmetry breaking, and two states $\alpha$ associated with the symmetry breaking induced by the Yukawa interactions.

All in all, the original symmetry is U(2) and breaks to U(1) if we include both symmetry breaking effects. The U(1) is a preserved rotational symmetry in the space transverse to the $\vec{\alpha}$ direction.

\section{Calculating the gravitational wave spectrum from the phase transition}\label{grav} 

A Spontaneous Symmetry Breaking (SSB) as the one proposed in section \ref{SSB} is a natural source of gravitational waves. Gravitational waves are in fact sourced by three processes, the expanding vacuum bubble collisions, sound waves and magnetohydrodynamic turbulences of the vacuum bubbles embedded in the hot plasma \cite{Kosowsky:1992vn}\cite{Kamionkowski:1993fg}\cite{Grojean:2006bp}\cite{Jaeckel:2016jlh}. Despite these being first order phase transition sources, it is stated that a second order SSB can generate this effects as well.\cite{Giblin:2011yh}

For the computation of these effects, we choose to put the potential in isovector notation and consider the couplings after renormalization (see Appendix). The potential then becomes:

\begin{equation}\label{zeroTpot}
V=m^2\left(\vec{\varphi}^*\vec{\varphi}\right)+\frac{\lambda_1}{2}\left(\vec{\varphi}^*\vec{\varphi}\right)^2-\frac{\lambda_2}{4}\left(\vec{\varphi}^*\vec{\varphi}^*\right)\left(\vec{\varphi}\vec{\varphi}\right)
\end{equation}

In this potential, the mass term is a parameter of the evolved theory (follow Appendix) and, since in equation \ref{bhl-pot} we have the negative of $M^2$ extracted (with a positive sign the squared mass would be negative as expected in a SSB), we will have that sign absorbed as $m^2\propto-M^2$. It must be clear that this is a potential compatible with all the symmetries of the initial sterile neutrino condensate potential we have developed through this article, with a particular choice of the parameter set to be used as a toy model. It is not meant to be the ultimate potential describing the sterile neutrino composite Higgs dynamics.

For our estimations we chose this squared mass related to the v.e.v. (equation \ref{veveq}) as $\left|m^2\right|\simeq10^{-2}\upsilon$. The couplings are selected in a way that the running is stable at every energy scale (figure \ref{fig:running} in Appendix) with values $\lambda_1=0.23$  and $\lambda_2=0.94$. This is just one amongst the many choices available for solving this problem, with no other singular interest. We have selected these particular values for illustration purposes exclusively.

\begin{figure}[H]\begin{center}
\captionsetup{width=0.8\textwidth}
\includegraphics[width=0.9\linewidth]{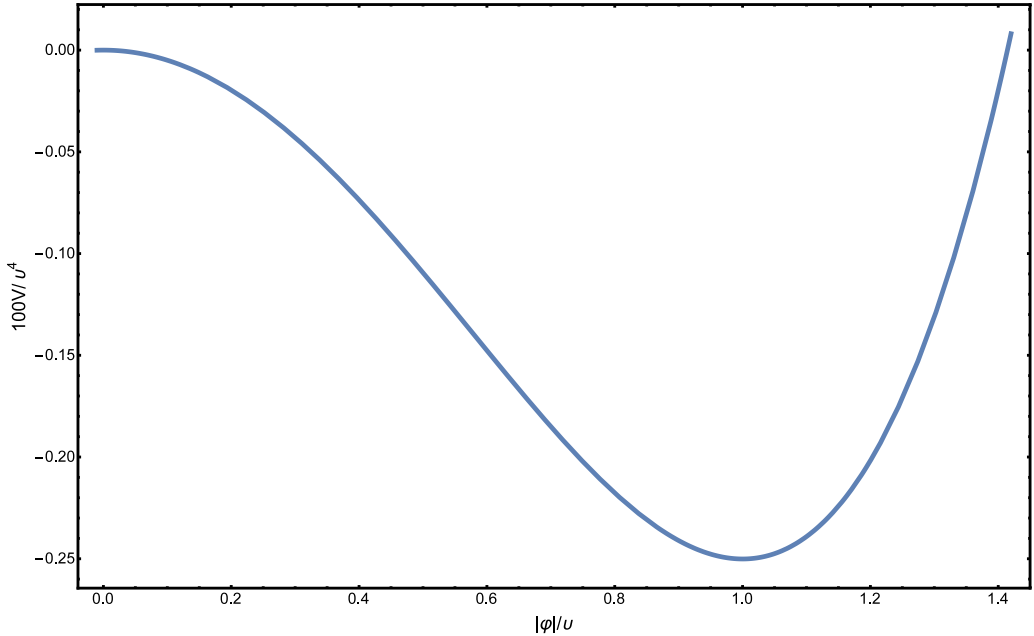}
\end{center}
\caption{\label{fig:VzeroT}Potential at zero temperature.}
\end{figure}

\begin{figure}[H]\begin{center}
\includegraphics[width=0.9\linewidth]{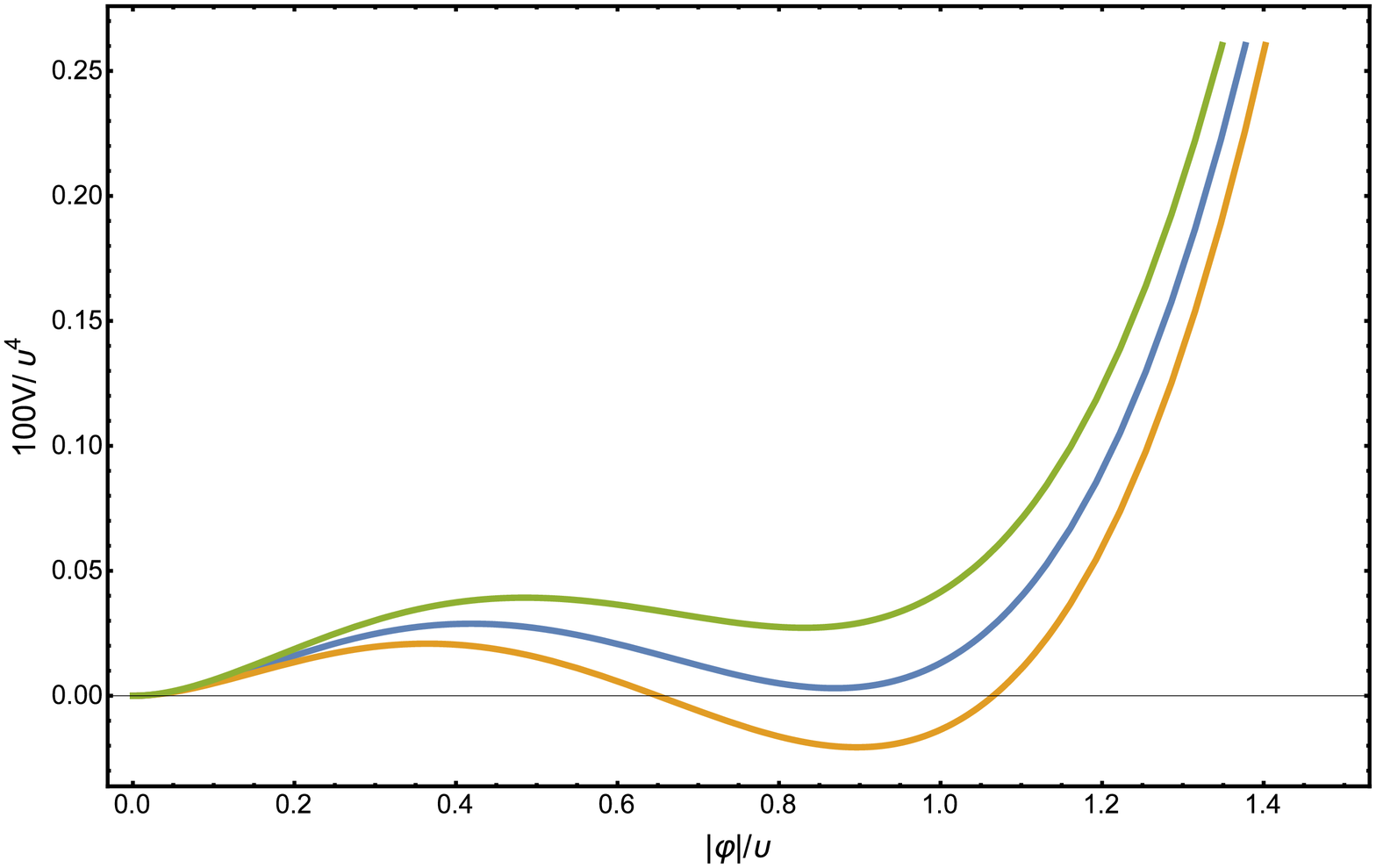}
\end{center}
\captionsetup{width=0.8\linewidth}
\caption{\label{fig:VT}Evolution of the potential according to temperature, using $V+\Delta V$ as in equation \ref{zeroTpot} and equation \ref{Tpot} for $T=0.31\upsilon$ (orange), $T=0.32\upsilon$ (blue) and $T=0.33\upsilon$ (green).}
\end{figure}

The SSB scale would naturally be much lower than the condensation scale, which makes it a second order phase transition. However, the difference between scales can be slightly reduced by fine tuning the value of $m^2$. For that reason, we will explore different values for $\upsilon$, and therefore, different values of $m^2$, at the end of this section. Although the standard line is to associate first order phase transitions with significant gravitational wave production \cite{Jaeckel:2016jlh}, the potential developed by the RH-neutrino condensate can source an interesting signal for an appropriate choice of parameters, as can be seen in figures \ref{fig:VzeroT} and \ref{fig:VT}.
	
Since the fields have two components, one of them complex, a path has been chosen for the analysis, defining them according to the parameter $t$ as $\vec{\varphi}=\left(t/\sqrt{2},0,t/2+i t/2\right)$

As we can see in figure \ref{fig:VzeroT}, the true vacuum of this potential at zero temperature is not located at $\left\langle\varphi\right\rangle=0$, in agreement to the vacuum of the potential this is inspired in, presented at section \ref{SSB}. At higher temperatures, the vacuum of the theory will be located at the origin, as shown in figure \ref{fig:VT}. This temperature dependence is introduced in the standard way, by adding the following term:

\begin{equation}\label{Tpot}\resizebox{\linewidth}{!}{$
\Delta V_T=\frac{3T^2}{2\pi^2}\left[\int^{\infty}_0 dq q^2\log{\left(1-\exp{\left(-\sqrt{q^2+\vec{\varphi}^*\vec{\varphi}\left(\frac{g(\mu)}{T}\right)^2}\right)}\right)}+\int^{\infty}_0 dq q^2\log{\left(1-\exp{\left(-q\right)}\right)}\right]
$}
\end{equation}

With two integrals accounting for the three massive bosons and the three massless ones \cite{Jaeckel:2016jlh}\cite{Dolan:1973qd}. The v.e.v. will be defined as $\upsilon=1$ through the whole calculation, leaving the factor accounting for its real value in the determination of the gravitational wave frequencies.

The thin wall approximation \cite{Jaeckel:2016jlh} has two markers of optimality. The first one is the gap between the true vacuum $ \left(\upsilon_{T_C}\right) $ at the critical temperature ($ T\simeq0.32\upsilon $ for us) and the energy at the origin:

\begin{equation}
\epsilon=V_{T_C}\left(0\right)-V_{T_C}\left(\upsilon_{T_C}\right)\simeq0.0698
\end{equation}

This value has to be $ 0<\epsilon\ll 1 $, which fits our model. The second reference for gravitational waves generation comes from the three-dimensional action, defined by:

\begin{equation}
S_3=4\pi\int^R_0 dr\,r^2V_T\left(\upsilon_{T_C}\right)+4\pi R^2\int^{\upsilon_{T_C}}_0d\varphi\sqrt{2V_T\left(\varphi\right)}
\end{equation}

Where R is the bubble radius. We choose $ T_*=0.29\upsilon $ as the temperature that makes an extreme action (with respect to the bubble radius) of $ S_3/T_*\simeq S_3/T_C\simeq105 $, which as explained in \cite{Jaeckel:2016jlh}, maximizes the probability of tunneling from the vacuum at $ \left\langle\varphi\right\rangle=0 $ to the true vacuum once the temperature has decreased.

After selecting an appropriate temperature for the transition to leave an interesting gravitational wave signal, we proceed to extracting the relevant parameters, which are $ \alpha $ and $ \beta $, defined as:

\begin{equation}
\begin{aligned}
&\alpha=\frac{\rho_{vac}}{\rho^*_{rad}}=\frac{\epsilon}{g_*\pi^2T^4_*/30}=0.97 \\ \\
&\frac{\beta}{H_*}=\left[T\frac{d}{dT}\left(\frac{S_3}{T}\right)\right]_{T\rightarrowtail T_*}=-1600 \\ \\
\end{aligned}
\end{equation}

Where $g_*=100$ is the number of relativistic degrees of freedom. These two parameters are everything needed to calculate the frequency spectrum of the phase transition. First, we define the peak frequencies of bubble walls collisions (col), sound waves in the plasma (sw) and magnetohydrodynamic turbulences after the collisions (mhd) as:

\begin{equation}
\begin{aligned}
&f_{col}=16.5\cdot10^{-6}\left(\frac{0.62}{1.8-0.1v+v^2}\right)\left|\frac{\beta}{H_*}\right|\left(\frac{T_*\left(\upsilon\cdot\mathrm{\,in\, GeV}\right)}{100}\right)\left(\frac{g_*}{100}\right)^{\frac{1}{6}}\,\, \mathrm{Hz}\\ \\
&f_{sw}=1.9\cdot10^{-5}\left(\frac{1}{v}\right)\left|\frac{\beta}{H_*}\right|\left(\frac{T_*\left(\upsilon\cdot\mathrm{\,in\, GeV}\right)}{100}\right)\left(\frac{g_*}{100}\right)^{\frac{1}{6}}\,\, \mathrm{Hz}\\ \\
&f_{mhd}=1.42f_{sw} \\ \\
\end{aligned}
\end{equation}

And then, the gravitational waves background gets the following contributions:

\begin{equation}
\Omega_{GW}\sim\Omega_{col}+\Omega_{sw}+\Omega_{mhd}
\end{equation}

Defined by:
\begin{equation}
\begin{aligned}
&h^2\Omega_{col}=1.67\cdot10^{-5}\left(\frac{H_*}{\beta}\right)^2\left(\frac{\kappa_{col}\alpha}{1+\alpha}\right)^2\left(\frac{100}{g_*}\right)^{\frac{1}{3}}\frac{0.11\mathrm{v}^3}{0.42+\mathrm{v}^2}\frac{3.8\left(f/f_{col}\right)^{2.8}}{1+2.8\left(f/f_{col}\right)^{3.8}} \\ \\
&h^2\Omega_{sw}=2.65\cdot10^{-6}\left|\frac{H_*}{\beta}\right|\left(\frac{\kappa_{sw}\alpha}{1+\alpha}\right)^2\left(\frac{100}{g_*}\right)^{\frac{1}{3}}\mathrm{v}\frac{f}{f_{sw}}\left(\frac{7}{4+3\left(f/f_{sw}\right)^2}\right)^\frac{7}{2} \\ \\
&h^2\Omega_{mhd}=3.35\cdot10^{-4}\left|\frac{H_*}{\beta}\right|\left(\frac{\kappa_{sw}\alpha}{1+\alpha}\right)^\frac{3}{2}\left(\frac{100}{g_*}\right)^{\frac{1}{3}}\mathrm{v}\frac{\left(f/f_{mhd}\right)^3}{\left(1+f/f_{mhd}\right)^\frac{11}{3}\left(1+8\pi f/h_{*}\right)} \\ \\
\end{aligned}
\end{equation}

\begin{figure}[H]\begin{center}
\includegraphics[width=0.95\linewidth]{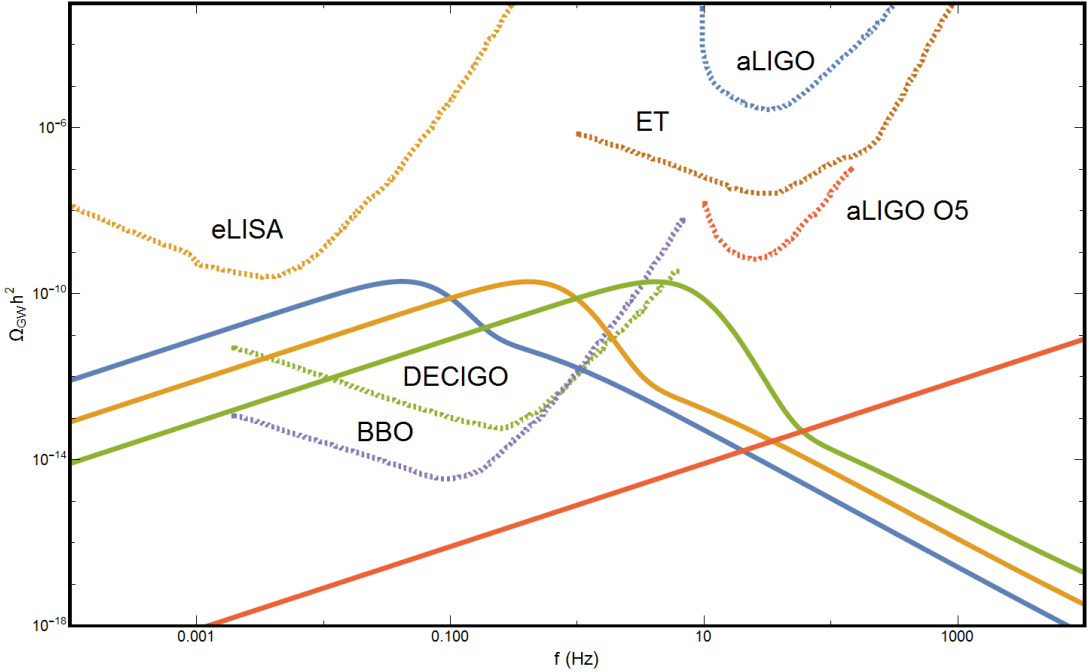}
\end{center}
\captionsetup{width=0.8\linewidth}
\caption{\label{fig:gravwaves}Gravitational wave signal for different mass and $\upsilon$ energy scales (1 TeV - Blue, 10 TeV - Orange, 100 TeV - Green, $10^{-9}\mathrm{M_P}$ - Red) compared to detectors reading spectra.}
\end{figure}

In this expressions, $\mathrm{v}$ is the bubble wall velocity (we have chosen $\mathrm{v}=1$) and the different $\kappa$ are the latent heat fractions (we decided to set all of them to $\kappa=0.5$). In figure \ref{fig:gravwaves}, the gravitational signal obtained with these equations is represented for different values of $\upsilon$.

\section{Conclusions}\label{conclusions}
Considering at least two generations of right-handed neutrinos has proven quite interesting for addressing several physical problems. First of all, we have determined, after successive breakings (spontaneous and explicit), that they leave composite scalars at three different mass scales that could be experimentally found and used as proof of the condensate formation and reveal information about it. Besides, these scalars could be of use for theories that require additional singlet degrees of freedom at high scales.

Inflation, already considered by \cite{Barenboim:2008ds}, would have a new set of particles being plausible inflaton candidates, a scenario probably worth considering. The composite scalar fields have a potential that may be interesting for an inflationary model as well.

The aid of these condensates for the stabilization of the Electroweak Vacuum has been an unfortunate failure, apparently due to having RGE equations with a weaker self-coupling than that presented in \cite{EliasMiro:2012ay}, thus not being enough for this purpose.

Besides, since we do not know yet whether the only scalar particle found in Nautre so far, the Higgs boson, is fundamental (see \cite{Strumia:2016wys}), this work presents a way to get a spectrum of scalar particles associated to different energy scales.

However, the implications of these neutrino condensates do not only end there. The gravitational waves signal hypothetically generated by the SSB of the their potential, estimated using toy model parameters, has been determined, obtaining a theoretical prediction that, if correct, would be measurable in the forthcoming years, when the experimental search of gravitational waves will presumably be a strong research field. Our results (figure \ref{fig:gravwaves}) show that a value from one to several tens up to few hundred TeV for $\Lambda$ (see section \ref{grav}) would be observable by DECIGO and BBO. Nonetheless, we would expect the bosons coming from the broken symmetries to show up in accelerator experiments too, putting bounds on the different energy scales present in the theory. A neutrino condensate located at a very high energy scale (figure \ref{fig:gravwaves}, in red) would not be observable by any current or planned detector.

\section*{Acknowledgments}
The authors want to acknowledge William A. Bardeen for his invaluable help and thorough guidance, which has made this project possible. We acknowledge support from the MEC and FEDER (EC) Grants SEV-2014-0398 and FPA2014-54459 and the Generalitat Valenciana under grant PROMETEOII/2013/017. This project has received funding from the European Unions Horizon 2020 research and innovation programme under the Marie Sklodowska-Curie grant Elusives ITN agreement No 674896 and In-visiblesPlus RISE, agreement No 690575. C.B. thanks \textit {Ministerio de Educaci\'on, Cultura y Deporte} for finantial support through an FPU-grant AP2010-3316. 

\section*{Appendix: Renormalization of the theory}\label{app} 

\subsection*{Majorana fermions:}

Let's stablish first how we jump from the usual Dirac notation to the one that best suits our study:
\begin{equation}
\gamma^\mu p_\mu \gamma^0=p^0-\gamma_5 \vec{\sigma}\vec{p}
\end{equation}

The RH-neutrino renormalized propagator will be (considering the contributions from the composite structure of the scalars) of the form:

\begin{equation}
\left\langle\psi_R\left(x\right)\psi^\dagger_R\left(x\right)\right\rangle_{\alpha\beta}=\frac{i}{\left(2\pi\right)^4}\int d^4p e^{-ip^\mu\left(x-y\right)_\mu}\frac{\left(p_0+\vec{\sigma}\vec{p}\right)_{\alpha\beta}}{p^2}
\end{equation}

Which, if we apply the same transformation we are using to change the notation into the isovector formalism:
\begin{equation}
\left(-i\sigma_2\right)_{\alpha\gamma}\left\langle\psi^{*}_R\left(x\right)\psi^T_R\left(x\right)\right\rangle_{\gamma\delta}\left(i\sigma_2\right)_{\delta\beta}=\frac{i}{\left(2\pi\right)^4}\int d^4pe^{-ip^\mu\left(x-y\right)_\mu}\frac{\left(p_0-\vec{\sigma}\vec{p}\right)_{\alpha\beta}}{p^2}
\end{equation}

We define the Majorana mass operators as:

\begin{equation}
\mathcal{O}_\psi\left(x\right)=\left(i\sigma_2\right)_{\alpha\beta}\psi_{R\alpha}\left(x\right)\psi_{R\beta}\left(x\right);\,\mathcal{O}^\dagger_\psi\left(x\right)=\left(-i\sigma_2\right)_{\alpha\beta}\psi^{*}_{R\alpha}\left(x\right)\psi^{*}_{R\beta}\left(x\right)
\end{equation}

That we will include in the fermion bubble function of neutrinos, in order to calculate this loop contribution:

\begin{equation}
\begin{aligned}
-i\int&d^4y e^{ik^\mu y_\mu}\left\langle O_\psi\left(x\right)O^\dagger_\psi\left(y\right)\right\rangle= \\ =&-i\int d^4y e^{ik^\mu y_\mu}\left\langle\left(i\sigma_2\right)_{\alpha\gamma}\psi_{R\alpha}\left(x\right)\psi_{R\beta}\left(x\right),\left(-i\sigma_2\right)_{\gamma\delta}\psi^*_{R\gamma}\left(y\right)\psi^*_{R\delta}\left(y\right)\right\rangle= \\
=& -i\int d^4y e^{ik^\mu y_\mu}\left(-2\right)\left\langle\psi_{R\beta}\left(x\right)\psi^*_{R\gamma}\left(y\right)\right\rangle\left(-i\sigma_2\right)_{\gamma\delta}\left\langle\psi^*_{R\delta}\left(y\right)\psi_{R\alpha}\left(x\right)\right\rangle\left(\sigma_2\right)_{\alpha\beta}= \\
=& \left(-2\right)\frac{i}{\left(2\pi\right)^4}\int d^4p\frac{\left(p_0+\vec{\sigma}\vec{p}\right)_{\beta\gamma}}{p^2}\frac{\left(\left(p+k\right)_0-\vec{\sigma}\left(\vec{p}+\vec{k}\right)\right)_{\gamma\beta}}{\left(p+k\right)^2}= \\
=& -2\frac{i}{\left(2\pi\right)^4}\int d^4p\frac{\mathrm{Tr}\left[\left(p_0+\vec{\sigma}\vec{p}\right)\left(\left(p+k\right)_0-\vec{\sigma}\left(\vec{p}+\vec{k}\right)\right)\right]}{p^2\left(p+k\right)^2}= \\
=&  -2\frac{i}{\left(2\pi\right)^4}\int d^4p\frac{2p_\mu\left(p+k\right)^\mu}{p^2\left(p+k\right)^2}= \\
=& -4\frac{i}{\left(2\pi\right)^4}\int d^4p\frac{1}{p^2}+2k^2\frac{i}{\left(2\pi\right)^4}\int	d^4p\frac{1}{p^2\left(p+k\right)^2} \\
\end{aligned}
\end{equation}

These mass operators will appear in the condensation mechanism, a BHL analysis equivalent to that peformed to the top quark in the literature:

\begin{equation}
\begin{aligned}
\mathcal{L}&=G\mathcal{O}_\psi\left(x\right)\mathcal{O}^\dagger_\psi\left(x\right)=-\tfrac{g^2}{G}\varphi^*\left(x\right)\varphi\left(x\right)-g\left[\varphi^*\mathcal{O}_\psi+\varphi\mathcal{O}^\dagger_\psi\right]\left(x\right) \\ 
&\rightarrow Z_\varphi\partial\varphi^*\partial\varphi-\Delta m^2\varphi^*\varphi -\tfrac{1}{2}\lambda\left(\varphi^*\varphi\right)^2
\end{aligned}
\end{equation}

\begin{equation}
Z_\varphi\left(\mu^2\right)=2g^2\frac{1}{\left(4\pi\right)^2}\mathrm{Ln}\frac{\Lambda^2}{\mu^2}
\end{equation}

\begin{equation}
\Delta m^2\left(\mu^2\right)=\frac{g^2}{G}-4g^2\frac{1}{\left(4\pi\right)^2}\Lambda^2
\end{equation}

Where we can see the renormalized mass and kinetic term below the scale $\Lambda$, which is a dummy variable that we will set as the scale where the RH-neutrinos have effectively condensed.

We can renormalize the four scalar coupling constant in the following way:

\begin{equation}
\begin{aligned}
\lambda\left(\mu^2\right)=&\frac{1}{2}g^4\left(-i\int d^4x\right)\left(-i\int d^4y\right)\left(-i\int d^4z\right)\left\langle \mathcal{O}_\psi\left(0\right)\mathcal{O}^\dagger_\psi\left(x\right)\mathcal{O}_\psi\left(y\right)\mathcal{O}^\dagger_\psi\left(z\right)\right\rangle= \\
=& -\frac{1}{2}g^4\frac{i}{\left(2\pi\right)^4}\int d^4p \,2\left(2\right)^3\frac{\mathrm{Tr}\left[\left(p_0+\vec{\sigma}\vec{p}\right)\left(p_0-\vec{\sigma}\vec{p}\right)\left(p_0+\vec{\sigma}\vec{p}\right)\left(p_0-\vec{\sigma}\vec{p}\right)\right]}{\left(p^2\right)^4}= \\
=& -16g^4\frac{i}{\left(2\pi\right)^4}\int d^4p\frac{1}{\left(p^2\right)^2}=16g^4\frac{1}{\left(4\pi\right)^2}\mathrm{Ln}\left(\frac{\Lambda^2}{\mu^2}\right)
\end{aligned}
\end{equation}

Which will leave the following effective couplings at the scale $\mu$ (BHL approximation).

\begin{equation}
g^2\rightarrow \frac{g^2}{Z_\varphi}=\frac{1}{2}\frac{\left(4\pi\right)^2}{\mathrm{Ln}\left(\Lambda^2/\mu^2\right)}
\end{equation}

\begin{equation}
\lambda\rightarrow \frac{\lambda}{Z^2_\varphi}=4\frac{\left(4\pi\right)^2}{\mathrm{Ln}\left(\Lambda^2/\mu^2\right)}
\end{equation}

\subsection*{Two flavour theory}

In this section we will study the dynamical effects of our effective theory of condensed scalars. Interpreting the symmetric tensor condensate field as an isovector field, the extension of the mass operators will be:

\begin{equation}
\mathcal{O}^{ij}_\psi\left(x\right)=\left(i\sigma_2\right)_{\alpha\beta}\psi^i_{R\alpha}\left(x\right)\psi^j_{R\beta}\left(x\right);\, O^{\dagger ij}_\psi\left(x\right)=\left(-i\sigma_2\right)_{\alpha\beta}\psi^{*i}_{R\alpha}\left(x\right)\psi^{*j}_{R\beta}\left(x\right)
\end{equation}

Leaving the following interaction Lagrangian:

\begin{equation}
\mathcal{L}_I=G\mathcal{O}^{ij}_\psi\left(x\right)\mathcal{O}^{\dagger ij}_\psi\left(x\right)=-\frac{g^2}{G}\varphi^{*ij}\varphi^{ij}\left(x\right)-g\left[\varphi^{*ij}\mathcal{O}^{ij}_\psi+\varphi^{ij}\mathcal{O}^{\dagger ij}_\psi\right]\left(x\right)
\end{equation}

Setting the same renormalization constants we have already presented:

\begin{equation}
\rightarrow Z_\varphi\partial\varphi^{*ij}\partial\varphi^{ij}-\Delta m^2\varphi^{*ij}\varphi^{ij}-\frac{\lambda}{2}\left(\varphi^{*ij}\varphi^{jk}\varphi^{*kl}\varphi^{li}\right)
\end{equation}

Using now the isovector Notation:

\begin{equation}
\varphi^{ij}\rightarrow\frac{1}{\sqrt{2}}\left(\vec{\tau}\vec{\varphi}i\tau_2\right)^{ij}
\end{equation}

\begin{equation}
\begin{aligned}
\Rightarrow& Z_\varphi\partial\varphi^{*ij}\partial\varphi{ij}-\Delta m^2\varphi^{*ij}\varphi^{ij}-\frac{\lambda}{2}\left(\varphi^{*ij}\varphi^{jk}\varphi^{*kl}\varphi^{li}\right)=\\
=&\frac{Z_\psi}{2}\mathrm{Tr}\left[\partial\left(\left(\vec{\tau}\vec{\varphi}^*\right)\partial\left(\vec{\tau}\vec{\varphi}\right)\right)\right]-\frac{\Delta m^2}{2}\mathrm{Tr}\left[\left(\vec{\tau}\vec{\varphi}^*\right)\left(\vec{\tau}\vec{\varphi}\right)\right]-\frac{\lambda}{8}\mathrm{Tr}\left[\left(\vec{\tau}\vec{\varphi}^*\right)\left(\vec{\tau}\vec{\varphi}\right)\left(\vec{\tau}\vec{\varphi}^*\right)\left(\vec{\tau}\vec{\varphi}\right)\right]=\\
=&Z_\psi\left(\partial\vec{\varphi}^*\partial\vec{\varphi}\right)-\Delta m^2\left(\vec{\varphi}^*\vec{\varphi}\right)-\frac{\lambda}{4}\left(2\left(\vec{\varphi}^*\vec{\varphi}\right)^2-\left(\vec{\varphi}^*\right)^2\left(\vec{\varphi}\right)^2\right)
\end{aligned}
\end{equation}

Where $Z_\varphi$, $\Delta m^2$ and $\lambda$ are the same as before:
\begin{equation}
Z_\varphi\left(\mu^2\right)=2g^2\frac{1}{\left(4\pi\right)^2}\mathrm{Ln}\left(\frac{\Lambda^2}{\mu^2}\right)
\end{equation}

\begin{equation}
\Delta m^2\left(\mu^2\right)=\frac{g^2}{G}-4g^2\frac{\Lambda^2}{\left(4\pi\right)^2}
\end{equation}

\begin{equation}
\lambda\left(\mu^2\right)=\frac{16g^4}{\left(4\pi\right)^2}\mathrm{Ln}\left(\frac{\Lambda^2}{\mu^2}\right)
\end{equation}

The computation of the bosonic loops gives (isovector normalization):

\begin{equation}
\mathcal{L}=\partial\vec{\varphi}^*\partial\vec{\varphi}-m^2\vec{\varphi}^*\vec{\varphi}-\frac{\lambda_1}{2}\left(\vec{\varphi}^*\vec{\varphi}\right)^2+\frac{\lambda_2}{4}\left(\vec{\varphi}^*\right)^2\left(\vec{\varphi}\right)^2
\end{equation}

\begin{equation}
\begin{aligned}
\Delta\left(\lambda_1\right)=&\left[3\lambda^2_1+2\lambda_1\lambda_2+\lambda^2_1\right]\frac{i}{\left(2\pi\right)^4}\int d^4p\frac{1}{\left(p^2\right)^2}=\\
=&-\left[4\lambda^2_1+2\lambda_1\lambda_2\right]\frac{1}{\left(4\pi\right)^2}\mathrm{Ln}\left(\frac{\Lambda^2}{\mu^2}\right)
\end{aligned}
\end{equation}

\begin{equation}
\begin{aligned}
\Delta\left(\lambda_2\right)=&\left[\frac{3}{2}\lambda^2_2+2\lambda_1\lambda_2\right]\frac{i}{\left(2\pi\right)^4}\int d^4p\frac{1}{\left(p^2\right)^2}=\\
=&-\left[\frac{3}{2}\lambda^2_2+2\lambda_1\lambda_2\right]\frac{1}{\left(4\pi\right)^2}\mathrm{Ln}\left(\frac{\Lambda^2}{\mu^2}\right)
\end{aligned}
\end{equation}

\begin{equation}
\Delta\left(\lambda_1\right)=-\lambda_1\left(4\lambda_1+2\lambda_2\right)\frac{1}{\left(4\pi\right)^2}\mathrm{Ln}\left(\frac{\Lambda^2}{\mu^2}\right)
\end{equation}

\begin{equation}
\Delta\left(\lambda_2\right)=-\lambda_2\left(\frac{3}{2}\lambda_2+2\lambda_1\right)\frac{1}{\left(4\pi\right)^2}\mathrm{Ln}\left(\frac{\Lambda^2}{\mu^2}\right)
\end{equation}

Now adding the fermion loop contribution (conventional definition with bosonic kinetic term normalized to 1).

\begin{figure}\begin{center}
\subfloat[]{\includegraphics[width=0.45\linewidth]{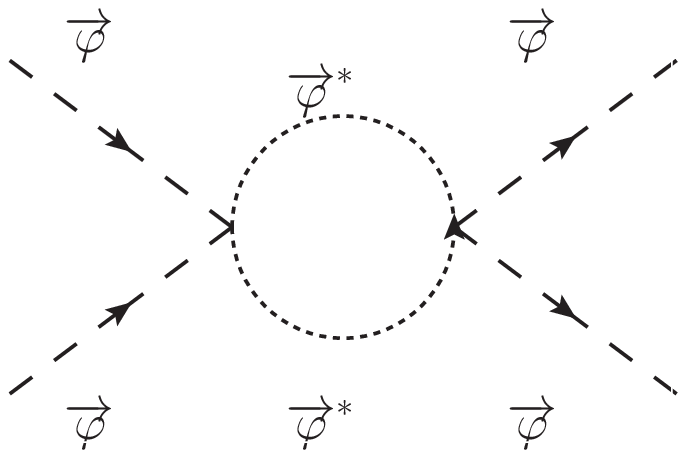}}\hspace{1cm}\subfloat[]{\includegraphics[width=0.45\linewidth]{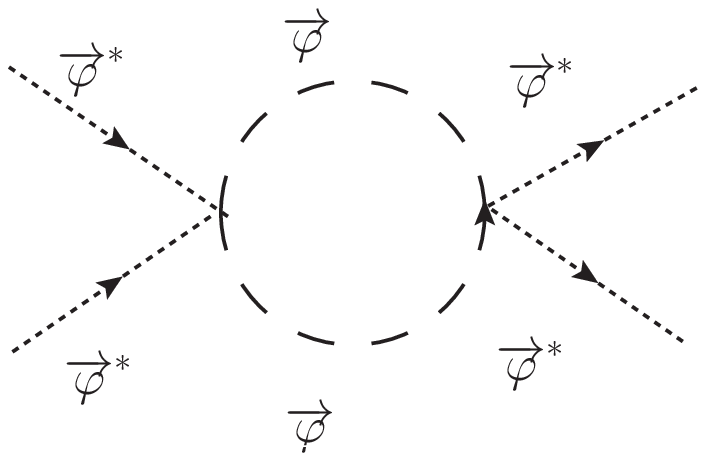}} \\ \hspace{0.9cm}\subfloat[]{\includegraphics[scale=0.9]{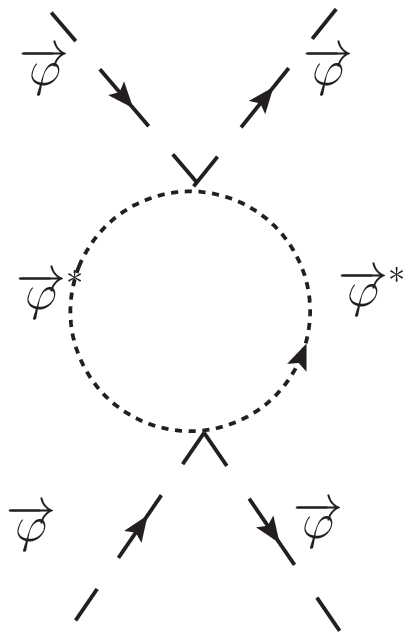}}\hspace{3.2cm}\subfloat[]{\includegraphics[scale=0.9]{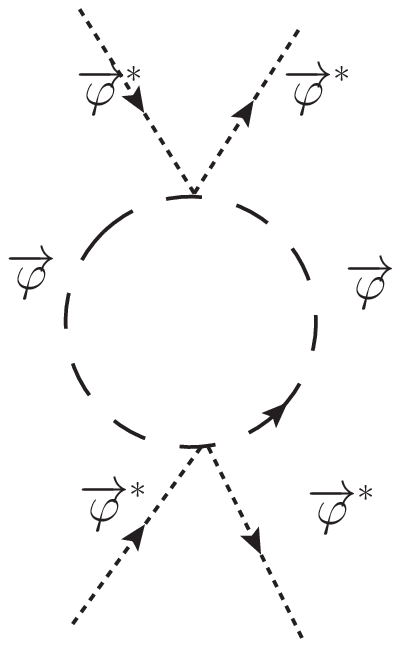}} \\ \subfloat[]{\includegraphics[width=0.45\linewidth]{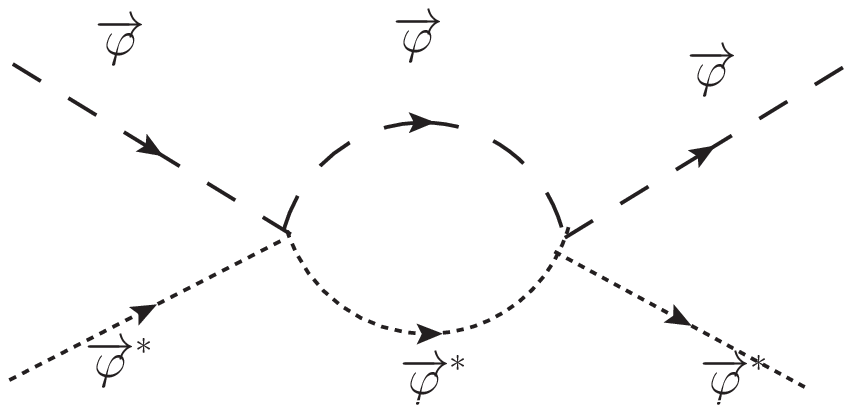}}\hspace{2.4cm}\subfloat[]{\includegraphics[scale=0.9]{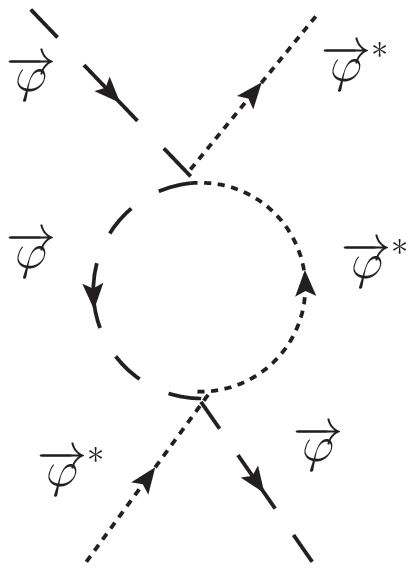}}\hspace{4.2cm}\end{center}\caption{\label{fig:loops2}Loops involved in the renormalization of $\lambda_1$ and $\lambda_2$.}
\end{figure}

\begin{equation}
\Delta\left(\lambda_1\right)=\left[-\lambda_1\left(4\lambda_1+2\lambda_2\right)-4g^2\lambda_1+16g^4\right]\frac{1}{\left(4\pi\right)^2}\mathrm{Ln}\left(\frac{\Lambda^2}{\mu^2}\right)
\end{equation}

\begin{equation}
\Delta\left(\lambda_2\right)=\left[-\lambda_2\left(\frac{3}{2}\lambda_2+2\lambda_1\right)-4g^2\lambda_2-16g^4\right]\frac{1}{\left(4\pi\right)^2}\mathrm{Ln}\left(\frac{\Lambda^2}{\mu^2}\right)
\end{equation}

The right-handed fermion mass matrix is:

\begin{equation}
\left(MM^\dagger\right)_{ij}=2g\varphi_{ik}2g\varphi^*_{kj}=2g^2\left[\left(\vec{\tau}\vec{\varphi}\right)\left(\vec{\tau}\vec{\varphi}^*\right)\right]_{ij}
\end{equation}

And the fermion wavefunction renormalization:

\begin{equation}
\begin{aligned}
&\frac{\left(p_0+\vec{\sigma}\vec{p}\right)}{p^2}\frac{\tau^ai\tau_2}{\sqrt{2}}\frac{i}{\left(2\pi\right)^4}\int d^4k\frac{1}{k^2}\frac{\left(\left(p+k\right)_0-\vec{\sigma}\left(\vec{p}+\vec{k}\right)\right)}{p^2}2g\frac{\left(-i\tau_2\right)\tau^a}{\sqrt{2}}\frac{\left(p_0+\vec{\sigma}\vec{p}\right)}{p^2}=\\
&=6g^2\frac{\left(p_0+\vec{\sigma}\vec{p}\right)}{p^2}\frac{i}{\left(2\pi\right)^4}\int d^4k\left(\left(p+k\right)_0-\vec{\sigma}\left(\vec{p}+\vec{k}\right)\right)\frac{\left(p_0+\vec{\sigma}\vec{p}\right)}{p^2}\\
&\cdot\int d\alpha\Gamma\left(2\right)\left[-\alpha\left(1-\alpha\right)p^2-\left(k+\alpha p\right)^2\right]^{-2}=\\
&=-\frac{6g^2}{\left(4\pi\right)^2}\frac{\left(p_0+\vec{\sigma}\vec{p}\right)}{p^2}\int d\alpha\left(1-\alpha\right)\left(p_0-\vec{\sigma}\vec{p}\right)\frac{p_0+\vec{\sigma}\vec{p}}{p^2}\Gamma\left(2-\tfrac{d}{2}\right)\left[-\alpha\left(1-\alpha\right)p^2\right]^{\tfrac{d}{2}-2}=\\
&=-\frac{3g^2}{\left(4\pi\right)^2}\frac{\left(p_0+\vec{\sigma}\vec{p}\right)}{p^2}\int d\alpha\mathrm{Ln}\left(\frac{\Lambda^2}{-\alpha\left(1-\alpha\right)p^2}\right)
\end{aligned}
\end{equation}

\begin{equation}
Z_\psi=1+\frac{3g^2}{\left(4\pi\right)^2}\mathrm{Ln}\left(\frac{\Lambda^2}{\mu^2}\right);\,\,Z_\varphi=1+\frac{2g^2}{\left(4\pi\right)^2}\mathrm{Ln}\left(\frac{\Lambda^2}{\mu^2}\right)
\end{equation}

The Yukawa renormalization is:

\begin{equation}
g\rightarrow \frac{g}{Z_\psi\sqrt{Z_\varphi}}=g-\frac{3g^3}{\left(4\pi\right)^2}\mathrm{Ln}\left(\frac{\Lambda^2}{\mu^2}\right)-\frac{g^2}{\left(4\pi\right)^2}\mathrm{Ln}\left(\frac{\Lambda^2}{\mu^2}\right)=g-\frac{4g^3}{\left(4\pi\right)^2}\mathrm{Ln}\left(\frac{\Lambda^2}{\mu^2}\right)
\end{equation}

The perturbative summary is thus made from the following equations:

\begin{equation}
\left(MM^\dagger\right)_{ij}=2g\varphi_{ik}2g\varphi^*_{kj}=2g^2\left[\left(\vec{\tau}\vec{\varphi}\right)\left(\vec{\tau}\vec{\varphi}^*\right)\right]_{ij}
\end{equation}

\begin{equation}
\Delta\left(g^2\right)=-\frac{8g^4}{\left(4\pi\right)^2}\mathrm{Ln}\left(\frac{\Lambda^2}{\mu^2}\right)
\end{equation}

\begin{equation}
\Delta\left(\lambda_1\right)=\left[-\lambda_1\left(4\lambda_1+2\lambda_2\right)-4g^2\lambda_1+16g^4\right]\frac{1}{\left(4\pi\right)^2}\mathrm{Ln}\left(\frac{\Lambda^2}{\mu^2}\right)
\end{equation}

\begin{equation}
\Delta\left(\lambda_2\right)=\left[-\lambda_2\left(\frac{3}{2}\lambda_2+2\lambda_1\right)-4g^2\lambda_1-16g^4\right]\frac{1}{\left(4\pi\right)^2}\mathrm{Ln}\left(\frac{\Lambda^2}{\mu^2}\right)
\end{equation}

And the one-loop perturbative beta-functions will be:

\begin{equation}
\beta_{g^2}=16\pi^2\left(\mu^2\partial\mu^2g^2\right)=8g^4
\end{equation}

\begin{equation}
\beta_{\lambda_1}=16\pi^2\left(\mu^2\partial\mu^2\lambda_1\right)=\left[\lambda_1\left(4\lambda_1+2\lambda_2\right)+4g^2\lambda_1-16g^4\right]
\end{equation}

\begin{equation}
\beta_{\lambda_2}=16\pi^2\left(\mu^2\partial\mu^2\lambda_2\right)=\left[\lambda_2\left(\frac{3}{2}\lambda_2+2\lambda_1\right)+4g^2\lambda_1+16g^4\right]
\end{equation}

Which have the potential for Coleman-Weinberg instability if $\left(2\lambda_1+\lambda_2\right)<0$ in IR.

BHL boundary conditions of the effective couplings at scale $\mu$ (bubble approximation with Landau poles at $\mu=\Lambda$).

BHL results:

\begin{equation}
g^2\rightarrow\frac{1}{2}\frac{\left(4\pi\right)^2}{\mathrm{Ln}\left(\Lambda^2/\mu^2\right)};\,\,\beta_{g^2}=16\pi^2\left(\mu^2\partial\mu^2g^2\right)=\frac{1}{2}\left(\frac{\left(4\pi\right)^2}{\mathrm{Ln}\left(\Lambda^2/\mu^2\right)}\right)^2
\end{equation}

\begin{equation}
\lambda_1\rightarrow4\frac{\left(4\pi\right)^2}{\mathrm{Ln}\left(\Lambda^2/\mu^2\right)};\,\,\beta_{\lambda_1}=16\pi^2\left(\mu^2\partial\mu^2\lambda_1\right)=4\left(\frac{\left(4\pi\right)^2}{\mathrm{Ln}\left(\Lambda^2/\mu^2\right)}\right)^2
\end{equation}

\begin{equation}
\lambda_2\rightarrow-4\frac{\left(4\pi\right)^2}{\mathrm{Ln}\left(\Lambda^2/\mu^2\right)};\,\,\beta_{\lambda_2}=16\pi^2\left(\mu^2\partial\mu^2\lambda_2\right)=-4\left(\frac{\left(4\pi\right)^2}{\mathrm{Ln}\left(\Lambda^2/\mu^2\right)}\right)^2
\end{equation}

Note: you would match the actual amplitudes at scale $\mu$, not the beta functions.

$\left[\lambda_i+\Delta\left(\lambda_i\right)\right]_{effective\;theory}=\left[\lambda_i\right]_{BHL}$ at scale $\mu$.

\begin{figure}[H]\begin{center}
\includegraphics[width=0.9\linewidth]{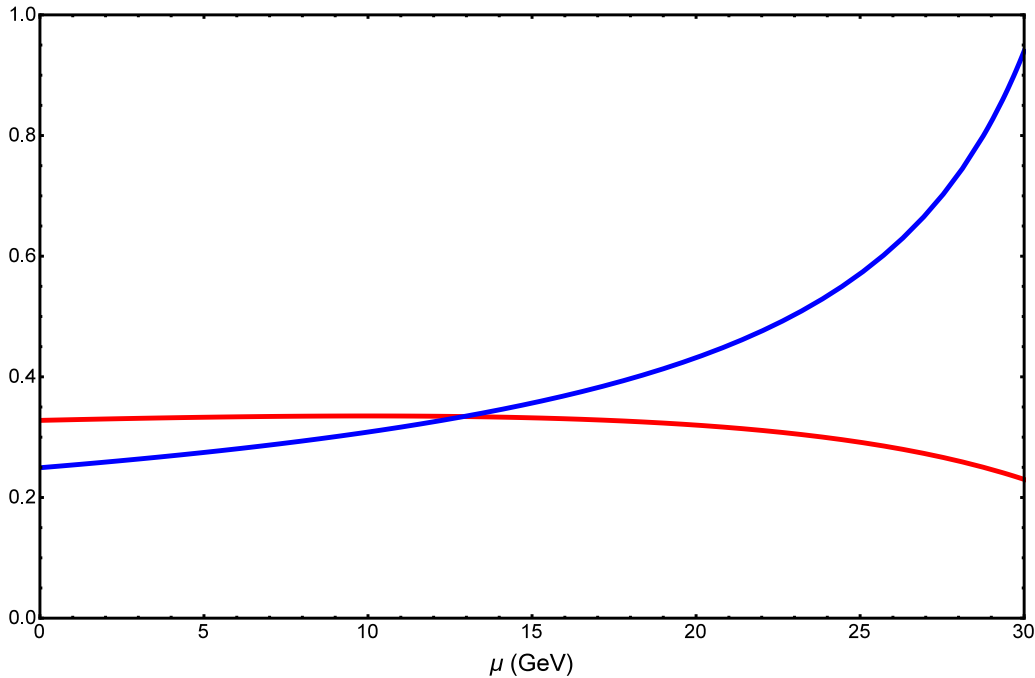}\end{center}
\captionsetup{width=0.8\linewidth}
\caption{\label{fig:running}Running of the couplings $\lambda_1$ (red) and $\lambda_2$ (blue) with boundaries selected for section \ref{grav}.}
\end{figure}

\end{document}